\documentclass[twocolumn,nofootinbib,preprintnumbers,amsmath,amssymb]{revtex4}

\usepackage{graphicx}
\usepackage{dcolumn}
\usepackage{bm}
\usepackage{amsmath}

\def \be {\begin{equation}}
\def \ee {\end{equation}}

\begin{document}

\title{Laboratory gamma-ray pulsar}

\author{Andrei Gruzinov}

\affiliation{ CCPP, Physics Department, New York University, 4 Washington Place, New York, NY 10003
}

\begin{abstract}

The mechanism by which gamma-ray pulsars shine might be reproducible in a laboratory. This claim is supported by three observations: (i) properly focusing a few PW optical laser gives an electromagnetic field in the so-called Aristotelian regime, when a test electron is radiation-overdamped; (ii) the Goldreich-Julian number density of this electromagnetic field (the number density of elementary charges needed for a nearly full conversion of optical power into gamma-rays) is of order the electron number density in a solid; (iii) above about $50$PW, the external source of electrons is not needed -- charges will be created by a pair production avalanche. 

\end{abstract}

\maketitle

\section{Introduction}

It appears that a gamma-ray pulsar can be created in a laboratory. Real pulsars are efficiently converting the large-scale Poynting flux into gamma-rays (up to order-unity efficiency for a weak axisymmetric pulsar, \cite{Gruzinov2014} and references therein). The laboratory pulsar is expected to efficiently convert optical light into gamma-rays.

At the level of estimates, the conditions needed for an efficient Poynting-to-gamma conversion appear to be reproducible in a laboratory. All one needs is (i) Aristotle number above one, meaning radiation damping stronger than inertia,  and (ii) the right (namely Goldreich-Julian, \cite{Goldreich1969}) number density of electrons. We discuss these two conditions in turn. 

\section{Aristotelian regime}

Consider a test electron in the electromagnetic field of generic geometry, $|E^2-B^2|\sim |{\bf E}\cdot {\bf B}|\sim F^2>0$, with characteristic length scale $\lambda$, and characteristic time scale $\lambda /c$. Let us estimate the characteristic Lorentz factor of the electron, $\gamma$. On the one hand, there exists a maximal possible $\gamma$ associated to the full ``potential drop'' of the field:
\be
\gamma _{\rm max}\sim {eF\lambda\over mc^2},
\ee
where $F\sim E\sim B$ is the characteristic value of electric and magnetic fields, $e$ is the electron charge, $m$ is the electron mass. On the other hand, there exists a terminal Lorentz factor at which the radiation damping balances the Lorentz force:
\be
\gamma _{\rm term}\sim \left( {F\lambda ^2\over e}\right)^{1/4}.
\ee
The electron is radiation-overdamped (the field is in the Aristotelian regime) if the terminal Lorentz factor is reached in less than the characteristic length scale, that is if
\be
\gamma _{\rm max}\gtrsim \gamma _{\rm term}.
\ee

Estimating the field from
\be
L\sim c\lambda^2F^2,
\ee
where $L$, erg/s, is the laser power, we write the condition of radiation overdamping as 
\be
{\rm Ar}\equiv {L\over L_e}\left({\lambda \over r_e}\right)^{-2/3}\gtrsim 1,
\ee
where we have defined the dimensionless Aristotle number Ar, with $L_e\equiv {mc^3\over r_e}=8.7\times 10^{16}$erg/s -- the classical electron luminosity, and $r_e\equiv {e^2\over mc^2}=2.8\times 10^{-13}$cm -- the classical electron radius.

Assuming that a (split) laser pulse of power $L_{\rm PW}\times 10^{22}$erg/s is focused onto a region of size $\lambda _{\mu}\times 10^{-4}$cm, we get an Aristotle number
\be
{\rm Ar}\sim 0.2L_{\rm PW}\lambda _{\mu}^{-2/3}.
\ee
For $\lambda _{\mu}=0.5$ and $L_{\rm PW}>3$, we have ${\rm Ar}>1$.

In Aristotelian regime, ${\rm Ar}\gtrsim 1$, the work done by the field goes into emission of curvature photons rather than into accelerating electrons. The characteristic photon energy is
\be
\epsilon \sim {mc^2\over \alpha}{\rm Ar}^{3/8},
\ee
where $\alpha$ is the fine structure constant. Pulsars have ${\rm Ar}\gg1$ and emit above about $1$GeV. ``Aristotelian lasers'' should emit above about $100$MeV.

\section{Goldreich-Julian number density}
Each electron emits gamma-rays at a power $\sim eFc$; if we want to convert the entire laser pulse into gamma-rays, the number density of electrons $n$ should be 
\be
n\sim {L\over \lambda^3eFc}\sim {c\lambda^2F^2\over \lambda^3eFc}\sim {F\over e\lambda}.
\ee
In pulsar physics, the last expression is known as the Goldreich-Julian density -- this is the number density of elementary charges needed to noticeably alter the external field $\sim F$. Numerically,
\be
n_{\rm GJ}\sim 1.2\times 10^{23}L_{\rm PW}^{1/2}\lambda _{\mu}^{-2}{\rm cm}^{-3}
\ee
is of order the number density in a solid.

We also note that above about $50$PW, the pair avalanche will (over) produce the necessary charge density starting from a single seed charge as described in \cite{Ruderman1975}: a seed charge emits gamma rays; gamma-rays pair produce in external magnetic field; pairs then emit gamma-rays, etc.

\section{Conclusion}

When powerful lasers are properly focused on a target or even on vacuum, an efficient optical to gamma-ray conversion should occur, enabled by the same mechanism by which the gamma-ray pulsars shine.


\begin{thebibliography}{99}

\bibitem{Gruzinov2014}
A. Gruzinov,  	arXiv:1402.1520 (2014)

\bibitem{Goldreich1969}
 P. Goldreich, W. H. Julian, Astrophys.\ J.\ {\bf 157}, 869 (1969)

\bibitem{Ruderman1975}
 M. A. Ruderman, P. G. Sutherland, Astrophys.\ J.\ {\bf 196}, 51 (1975)


\end{thebibliography}
\end{document}